\documentclass[%
 preprint,
 amsmath,amssymb,
 aps,
]{revtex4-1}

\usepackage{graphicx}
\usepackage{dcolumn}
\usepackage{bm}

\begin{document}

\preprint{APS/123-QED}

\title{Strain-controlled spin splitting in the conduction band of monolayer $WS_2$}

\author{Moh. Adhib Ulil Absor}
\email{adib@ugm.ac.id} 
\affiliation{Department of Physics Universitas Gadjah Mada BLS 21 Yogyakarta Indonesia.}%
\author{Hiroki Kotaka}
\affiliation{ISIR-SANKEN, Osaka University, Ibaraki, Osaka 567-0047, Japan.}%

\author{Fumiyuki Ishii}%
\affiliation{Faculty of Mathematics and Physics Institute of Science and Engineering Kanazawa University Kanazawa 920-1192 Japan.}%

\author{Mineo Saito}
\affiliation{Faculty of Mathematics and Physics Institute of Science and Engineering Kanazawa University Kanazawa 920-1192 Japan.}%

\date{\today}

\begin{abstract}
Spin splitting bands that arises in conduction band minimum (CBM) of $WS_2$ monolayer (ML) play an important role in the new spin-orbit phenomena such as spin-valley coupled electronics. However, application of strain strongly modifies electronic properties of the $WS_2$ ML, which is expected to significantly affect to the properties of the spin splitting bands. Here, by using fully-relativistic first-principles calculations based on density-functional theory, we show that a substantial spin spliting bands observed in the CBM is effectively controlled and tuned by applying the biaxial strain. We also find that these spin splitting bands induce spin textures exhibiting fully out-of-plane spin polarization in the opposite direction between the $K$ and $Q$ points and their time reversals in the first Brillouin zone. Our study clarify that the strain plays an significant role in the spin-orbit coupling of the $WS_2$ ML, which has very important implications in designing future spintronics devices. 
\end{abstract}

\pacs{Valid PACS appear here}
\maketitle

\section{INTRODUCTION}
Spin-orbit coupled systems that plays an important role in spintronics device operations have attracted considerable scientific interest over recent years. This spin-orbit coupling (SOC) allows for the manipulation of electron spin \cite {Kato}, leading to the interesting effect such as current-induced spin polarization \cite {Kuhlen}, the spin Hall Effect \cite {Qi1}, the spin galvanic-effect \cite {Ganichev}, and the spin ballistic transport \cite {Lu}. The electric tunability of SOC has also been achieved by using gated semiconductor heterostructures \cite {Nitta}, thereby opening a new gateway to applications ranging from spintronics to quantum computing. For instant, some of the various spintronics devices that have already been studied include spin-field effect transistors \cite {Datta}, spin filters \cite {Koga}, and spin qubit gates \cite {Foldi}.

One of the promising materials candidate for spintronics comes from monolayer (ML) transition metal dichalcogenides (TMDs) family because of their extraordenary properties such as spin-valley coupled electronic structructures \cite {Xu,Yuan, Bromley,Zhu,Latzke}. Especially $WS_2$ ML system attracted much attentions since its carrier mobility has been predicted to be the largest among the ML TMDs family \cite {Ovchinnikov}. Furthermore, energy bands of $WS_2$ ML have well separated valleys, together with strong SOC in the 5$d$ orbitlas of $W$ atoms, a large spin splitting [ 0.426 to 0.433 eV ] has been established \cite {Zhu,Latzke,Liu_Bin}. This large spin splitting which mainly apparents at the $K$ point in the valence band maximum (VBM) is believed to be responsible for inducing some of interesting phenomena such as spin Hall effect, spin-dependent selection rule for optical transitions, and magneto-electric effect in TMDs \cite {Gong}.	
	
Besides the well-studied of the spin splitting in the VBM, the spin splitting of the conduction band minimum (CBM) in the $WS_2$ ML also attracted much attentions since $n$-type system has been experimentally observed \cite {Ovchinnikov,Morrish}. In the CBM, there is two local minima exhibiting a completely different features of spin splitting. The first local minima, which is located on the $K$ point, has a substantially small spin splitting [ 26 to 29 meV ] \cite {Kosminder,Liu_Bin}. On the other hand, another local minima, which is located on the point approximately midway between the $\Gamma$ and $K$ point, namely $Q$ point, has large spin splitting \cite {Liu}, which is comparable to that of the $K$ point in the VBM \cite {Zhu,Latzke,Liu_Bin,Liu}. Because energy minimum at the $Q$ point is close to that of $K$ point in the CBM, the spin splitting bands in the $Q$ point is expected to play a significant role in the new spintronics properties such as spin-conserving scattering \cite {Liu}. However, the CBM of $WS_2$ ML is sensitively modified by application of strain \cite {David}. Therefore, it is important to clarify the effect of the strain on the spin-splitting of CBM, which is expected to induces useful electronic properties for spintronics devices.

In this paper, we performed first-principles density-functional calculation to clarify the spin splitting of the CBM on strained $WS_2$ ML system. We find that substantial spin spliting bands are identified in the CBM, which is effectively controlled and tuned by applying the strain. We also find that these spin spliting bands induces spin textures exhibiting fully out-of-plane spin polarization in the opposite directions between the $K$ and $Q$ points and their time reversals. We clarify the origin of these spin-split bands and spin textures by using simplified spin-orbit Hamiltonian derived from the group theory combined with the orbitals hybridization analyses. Finally, a possible application of the presence systems for spintronics will be discussed.

\section{Computational Methods}

Bulk monolayer $WS_2$ has stable polytype layered structures known as hexagonal ($2H$) structures. This structures has $P63/mmc$ ($D^4_{6h}$) space group, which consists of $S$-$W$-$S$ slabs weakly bounded by van der Walls interaction \cite {Bromley}. In such a slab, hexagonally layers of the $W$ atoms is sandwiched between two layers of the $S$ atoms through strong ionic-covalent bonds forming a trigonal prismatic arrangement [Fig. 1(a)]. Therefore, it is possible to create a stable $WS_2$ monolayer (ML) from the $2H$-$WS_2$ compounds by using micromechanical cleavage and liquid exfoliation \cite {Lee,Coleman}. Here, the trigonal prismatic coordination of the bulk $WS_2$ remains in the $WS_2$ ML, but its symmetry reduces to be $P6m2$ ($D^1_{3h}$) due to the lack of inversion symmetry \cite {Zhu}.

We performed first-principles electronic structure calculations on the $WS_2$ ML based on the density functional theory (DFT) within the generalized gradient approximation (GGA) \cite {Perdew} using the OpenMX code \cite{Openmx}.  We used norm-conserving pseudo-potentials \cite {Troullier}, and the wave functions are expanded by the linear combination of multiple pseudoatomic orbitals (LCPAOs) generated using a confinement scheme \cite{Ozaki,Ozakikino}. The orbitals are specified by $W$7.0-$s^{2}p^{2}d^{1}$ and $S$9.0-$s^{1}p^{1}d^{1}$, which means that the cutoff radii are 7.0 and 9.0 bohr for the $W$ and $S$ atoms, respectively, in the confinement scheme \cite{Ozaki,Ozakikino}. For the $W$ atoms, two primitive orbitals expand the $s$ and $p$ orbitals, and one primitive orbital expands the $d$ orbital. On the other hand, for the $S$ atoms, one primitive orbital expands the $s$, $p$, and $d$ orbitals. A 12x12x12 $k$-point grid is used. SOC was included in these fully relativistic calculations, and the spin textures in $k$-space were calculated using the $k$-space spin density matrix of the spinor wave function \cite{Kotaka, Absor1,Absor2,Absor3}.

The two dimensional structures of $WS_2$ ML are modelled as a periodic slab with a sufficiently large vacuum layer (25 \AA) in order to avoid interaction between adjacent layers. The geometries were fully relaxed until the force acting on each atom was less than 1 meV/\AA. We find that the optimized in-plane lattice constant of $WS_2$ ML is 3.18 \AA, which is in a good agreement with recently calculated reported data \cite{Zhu,David}. We consider a wide range of biaxial strains (up to 8\%), which is applied to the in-plane lattice constant. We define the degree of in-plane biaxial strain as $\epsilon_{xx}=\epsilon_{yy}=(a-a_0)/a_0$, where $a_0$ is the unstrained in-plane lattice constant. Here, we studied the following two different cases: the tensile strain, which increases the in-plane lattice constant $a$, and compressive strain, which decreases $a$.

\begin{figure}
	\centering
		\includegraphics[width=0.7\textwidth]{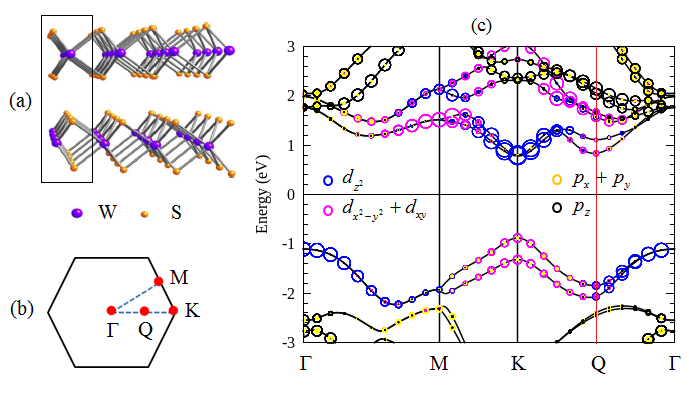}
	\caption{(a) Side view of $2H$-$WS_2$ unit cell, consisting of two $WS_2$ ML slab. (b) First Brillouin zone of monolayer $WS_2$ which is spesified by the high symmetry points ( $\Gamma$, $K$, $M$, and $Q$ points ). Here, the $Q$ point is defined as a point located approximately midway between the $\Gamma$ and $K$ point. (c) Orbital-resolved of the electronic band structures of $WS_2$ ML. The radius of circles reflects the magnitudes of spectral weight of the specific orbitals to the band.}
	\label{figure:Figure1}
\end{figure}

\section{RESULT AND DISCUSSION}

Fig. 1 (c) shows orbital-resolved of electronic band structures of the $WS_2$ ML calculated along the first Brillouin zone [Fig. 1(b)]. We find that, the VBM has two local maxima located on the $K$ and $\Gamma$ points, while the CBM has two local minima located on the $K$ and $Q$ points [Fig. 2(b)]. In the VBM, the local maxima at the $K$ point is dominated by $W$ $d_{{x^2}-{y^2}}+d_{xy}$ bonding states, while the local maxima at the $\Gamma$ point is predominately filled by $W$ $d_{z^2}$ bonding states. On the other hand, in the CBM, the local minima at the $K$ point is mainly derived from $W$ $d_{z^2}$ anti-bonding states and the local minima at the $Q$ point mainly originates from $W$ $d_{{x^2}-{y^2}}$ anti-bonding states. Here, we find that a direct band gap is clearly visible in the band structures where the topmost and lowest energy of the VBM and CBM, respectively, are observed in the $K$ point, which is consistent well with previous calculational results \cite {Zhu,Kosminder}.

By introducing the SOC, a substantial spin splitting of the band structures is established due to the lack of inversion symmetry. Here, spin degeneracy of the electronic band structures is broken, except for the $\Gamma-M$ lines due to time reversability [Figs. 2(a)-(c)]. In the case of equilibrium system, we find large spin splitting of 0.43 eV at the $K$ point in the VBM, which is in a good agreement with recently calculated reported data [ 0.426 to 0.433 eV ] \cite {Zhu,Latzke,Liu_Bin}. This large spin splitting mainly originates from the contribution of hybridization between  the $W$ $d_{{x^2}-{y^2}}+d_{xy}$  and $S$ ${p_x}+{p_y}$ bonding states. Furthermore, due to same contribution of the hybridization states, large spin splitting up to 0.33 eV is observed on the $Q$ point in the CBM, which is comparable with those of the $K$ point in the VBM. However, substantially small spin-splitting of 30 meV is found at the $K$ point in the CBM, which is due to the fact that the hybridization between $W$ $d_{z^2}$ and $S$ $p_z$ anti-bonding states contributes only small to the spin splitting \cite {Zhu}. This small value of spin splitting at the $K$ point in the CBM is consistent with previous calculated results obtained by the GGA \cite {Kosminder} as well as the tight-binding calculations \cite {Liu_Bin}. Because the energy minimum at the $Q$ point is close to that of the $K$ point in the CBM, it is expected that the spin splitting bands at the $Q$ point play a significant role in the new spintronics phenomena such as spin-conserving scattering. This is supported by the fact that a quantum interference due to the spin-conserving scattering processes involving the spin-splitting bands at the $Q$ point in the CBM has been experimentally observed in the $WSe_2$ ML \cite {Liu}.

\begin{figure}
	\centering
		\includegraphics[width=1.05\textwidth]{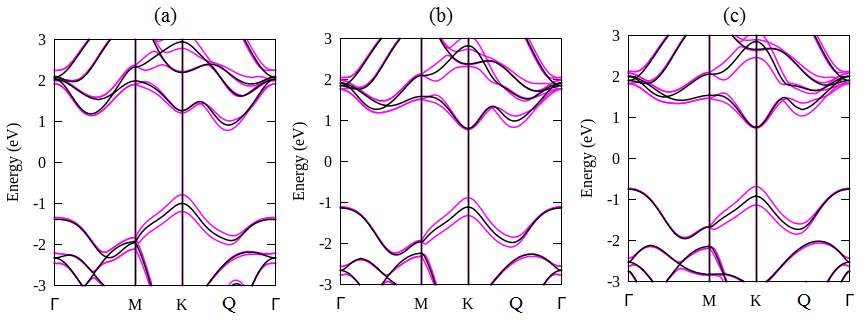}
	\caption{Electronic band structures of the strained $WS_2$ ML: (a) $\epsilon=-2\%$ (b) $\epsilon=0\%$, and (c) $\epsilon=2\%$. The black lines  (pink lines) indicates the band structures without (with) spin-orbit interaction. }
	\label{figure:Figure2}
\end{figure}

The application of strain on the $WS_2$ ML subsequently induces strong modification of its electronic band structures. As shown in Figs. 2(a)-(c), a transition from a direct to an indirect band gap is achieved when compressive or tensile strains are introduced. Under tensile strain, the $S$-$W$ bondlength ($d_{S-W}$) enhances and the $S$-$W$-$S$ angle ($\theta_{S-W-S}$) becomes small [Fig. 3(a)] and, consequently, hybridization between the $W$ $d_{z^2}$ and $S$ $p_z$ bonding states is reduced. On the other hand, the hybridization between the $W$ $d_{{x^2}-{y^2}}+d_{xy}$ and $S$ ${p_x}+{p_y}$ bonding states is strengthened. The increased hybridization between inplane bonding states [$W$ $d_{{x^2}-{y^2}}+d_{xy}$ and $S$ ${p_x}+{p_y}$] with tension is responsible for decreasing in energy of the $K$ point with respect to those of the $\Gamma$ point in the VBM [Fig. 3(b)]. Consistent with this argument, compressive strain leads to the fact that the energy level of the $W$ $d_{{x^2}-{y^2}}$ anti-bonding state shifts to be lower than that of the $W$ $d_{z^2}$ anti-bonding state, leading to the fact that the energy level of the $Q$ point becomes lower than that of the $K$ point in the CBM [Fig. 3(b)]. The shift in energy of the VBM and CBM induces electronic transition of the $WS_2$ ML from a direct to an indirect band gap \cite{David}, which is in fact confirmed by our calculated results [Figs. 2(a)-(c)].

\begin{figure}
	\centering
		\includegraphics[width=0.55\textwidth]{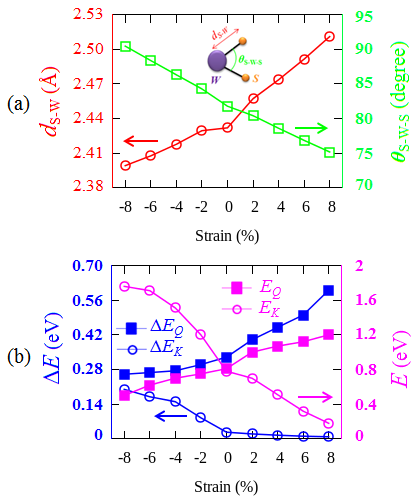}
	\caption{Relation between the structural parameters ( bondlength of the $S$-$W$ atoms, $d_{S-W}$, and the $S$-$W$-$S$ angle, $\theta_{S-W-S}$ ), energy minimum at the $K$ and $Q$ points ( $E_{K}$, $E_{Q}$ ) in the conduction band minimum (CBM), and energy of spin splitting bands of the CBM in the $K$ and $Q$ points ( $\Delta E_{K}$, $\Delta E_{Q}$ ) as a function of strain. (a) The bondlength $d_{S-W}$ (red circle-lines) and angle $\theta_{S-W-S}$ ( green sequare-lines) are given. (b) The energy minimum of the CBM (pink colours) calculated around the the $K$ point ($E_{K}$, circle-lines) and the $Q$ point ($ E_{Q}$, solid sequare-lines) and the energy of the spin splitting bands of the CBM (blue colours) calculated around the $K$ point ($\Delta E_{K}$, circle-lines) and the $Q$ point ($\Delta E_{Q}$, solid sequare-lines) are shown. }
	\label{figure:Figure3}
\end{figure}

Since the biaxial strain strongly modifies the electronic properties of the $WS_2$ ML, it is expected that a significant change of the spin splitting bands is achieved. Here, we focused on the spin splitting of the CBM because the $n$-type system is really achieved \cite {Ovchinnikov,Morrish}, which is supported by the fact that various the $n$-type systems such as $WSe_2$ \cite {W_Liu,Chuang} and $M_{0}S_{2}$ \cite {Ghatak,Yijin} ML systems has been experimentally observed. We find that, due to the increased overlap of the $W$ $d_{{x^2}-{y^2}}+d_{xy}$ and $S$ ${p_x}+{p_y}$ anti-bonding states, strong enhancement of the spin splitting bands in the $Q$ point is achieved under tensile strain [Fig. 3(b)]. However, it does'nt affect to the spin-splitting in the $K$ point due to the domination of the $W$ $d_{z^2}$ anti-bonding state. On the other hand, the coupling between the $W$ $d_{{x^2}-{y^2}}+d_{xy}$ and $S$ ${p_x}+{p_y}$ anti-bonding states is reduced when compressive strain is introduced, leading to the fact that the spin splitting bands reduce and enhance in the $Q$ and $K$ points, respectively. Therefore, it is concluded that the strain sensitively affects to the spin-splitting of the CBM in the $WS_2$ ML.

In order to better understand the nature of the observed spin splitting, we show in Fig. 4 the calculated results of the spin textures on strained $WS_2$ ML. By assuming that Fermi level is located on 225 meV above the CBM, we find that six-fold symmetry ( $C_{6}$ ) of spin-split Fermi pockets is observed on the equilibrium as well as the strained systems. In the case of equilibrium system, these Fermi pockets are clearly visible around the $K$ and $Q$ points exhibiting fully out-of-plane spin polarization in the opposite direction between the $K$ and $Q$ points and their time reversals [Fig. 4(b)]. These alternating directions of the spin polarization makes the symmetry of these Fermi pockets reduce to be three-fold ( $C_{3}$ ). Here, a similar features of the spin-split Fermi pockets has recently been reported in $WSe_2$ ML \cite {Liu}, indicating that our calculational results of the spin-splitt Fermi Pockets are consistent well with general properties of spin textures in the ML TMDs materials \cite {Liu,Yuan,Zhu}. Introducing strain subsequently modifies these Fermi pockets because of the shifting in energy of the CBM. In the case of tensile strain, the energy level of the $Q$ point becomes higher than that of the $K$ point, leading to the fact that only the $K$ spin-split Fermi pockets are observed [Fig. 4(c)]. On the other hand, introducing compressive strain subsequently shifts the $Q$ point to be lower energy than that of the $K$ point, resulting that only the $Q$ spin-split Fermi pockets are visible [Fig. 4(a)]. This considerably change of the spin-split Fermi pockets is expected to play a significant role in the spin-conserving scattering between the $Q$ and $K$ points, which is possible to implay the long spin lifetime and the long valley lifetime \cite {Liu}. In fact, long-lived nanosecond spin relaxation and spin coherence of electrons in the $WS_2$ ML has recently been reported \cite {L_Yang}.

To clarify the exsistence of the spin splitting and spin textures in our calculated results, we consider our system based on the symmetry arguments. The $WS_2$ ML system belongs to the the symmetry point group of $D_{3h}$. This symmetry itself combines the $C_{3v}$ symmetry group and a mirror reflection $M$ with respect to the hexagonal plane of the Brillouin zone [Fig. 1(a)-(b)]. In the case of two-dimensional system with $C_{3v}$ symmetry, the SOI Hamiltonian up to cubic $k$-terms can be expressed as \cite {Zhu, Yuan} 
\begin{equation}
\label{1}
H(k)=\alpha_{R}(k_{x}\sigma_{y}-k_{y}\sigma_{x})+\lambda_{k}(3{k^2_{x}}-{k^2_{y}})k_{y}\sigma_{z},
\end{equation}
where $k=\sqrt{{k^2_{x}}+{k^2_{y}}}$ and $\sigma_{i}$ are Pauli. Here, the first term in the $H(k)$ is the Rashba term characterized by Rashba parameter, $\alpha_{R}$, inducing in-plane spin polarizations. On the other hand, the second term in the $H(k)$ is the warping term characterized by warping parameter, $\lambda_{k}$, which contributes to the out-of-plane spin polarization. More importantly, the Rashba parameter $\alpha_{R}$ is induced by the out-of-plane potential gradient asymmetry and it is more sensitive to the hybridization between out-of-plane orbitals [ $W$ $d_{z^2}$ and $S$ $p_z$ ], while the warping parameter $\lambda_{k}$ is mainly contributed from in-plane potential gradient asymmetry, which is strongly affected by the hybridization between in-plane orbitals [ $W$ $d_{{x^2}-{y^2}}+d_{xy}$ and $S$ ${p_x}+{p_y}$ ] \cite {Vajna,Zhu}. Therefore, both the $\alpha_{R}$ and $\lambda_{k}$ should be sensitive to the application of the in-plane and out-of-plane strains.  The additional $M$ symmetry operation of $D_{3h}$ symmetry suppresses the Rashba term in the $H(k)$, leading to the fact that only the second term of the $H(k)$ remains. Here, the spin splitting and spin polarization are expressed as 
\begin{equation}
\label{2}
\Delta E(k,\theta)=\lambda_{k}\left|\sin(3\theta)\right|,
\end{equation}
and
\begin{equation}
\label{3}
P_{\pm}(k,\theta)=[0,0,\mp \lambda_{k}sin(3\theta)],
\end{equation} 
respectively, where $\theta =tan^{-1}(k_{y}/k_{x})$. Since the spin splitting depends only on the warping parameter $\lambda_{k}$ according to the Eq. (\ref{2}), it is expected that the magnitude of spin splitting can be tuned by applying in-plane strain.

\begin{figure}
	\centering
		\includegraphics[width=1.0\textwidth]{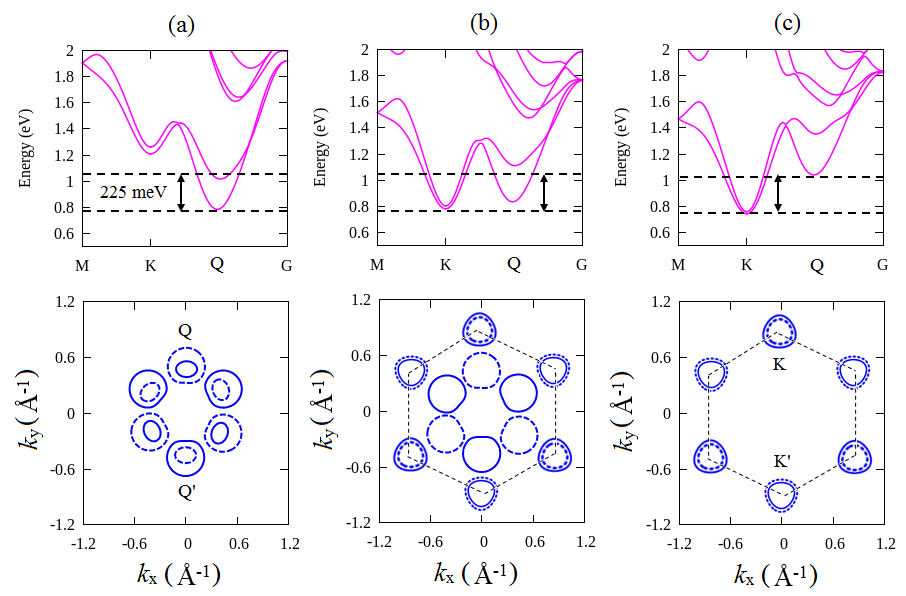}
	\caption{band structures of the CBM corresponding to the spin textures: (a) $\epsilon=-2\%$ (b) $\epsilon=0\%$, and (c) $\epsilon=2\%$. In the spin textures, the solid (dotted) lines represents up (down) spin states with out-of-plane orientations. The spin textures are calculated on the energy band of 225 meV above the CBM. }
	\label{figure:Figure4}
\end{figure}

As mentioned before that the application of strain sensitively affects to the structural parameters [Fig. 3(a)]. However, in the present calculations, the symmetry of the optimized structures is found to be invariant, leading to the fact that the expression of spin splitting and spin polarization in Eqs. (\ref{2}) and (\ref{3}) remains unchanged. Because the structural parameters ($d_{S-W}$ and $\theta_{S-W-S}$) significantly affect the coupling between the in-plane orbitals [ $W$ $d_{{x^2}-{y^2}}+d_{xy}$ and $S$ ${p_x}+{p_y}$ ], it is expected that considerably change of the magnitude of the warping parameter $\lambda_{k}$ is established. Consequently, a substantial change of the spin splitting in the electronic band structures is achieved, which is in fact reflected by our first-principles results [ Fig. 3(b) ]. Furthermore, considering the fact that these spin spliting bands exhibit $C_{3}$ symmetry of the spin-split Fermi pockets due to the alternating orientation of out-of-plane spin polarization between the $K$ and $Q$ points and their time reversals [ Fig. 4 (a)-(c)], this is also consistent with the $\sin 3\theta$ dependence of the spin polarization defined in the Eq. (\ref{3}). Therefore, it can be concluded that the calculated results of the spin splitting bands and spin textures are agree well with above-mentioned simplified Hamiltonian.
 
Thus far, we found that the spin-splitting of CBM on $WS_2$ ML can be controlled by applying biaxial strain. This strain can be achieved by introducing a lattice mismatch between $WS_2$ ML and the substrate or by applying doping atom. Recently, substrate induces giant spin splitting has been predicted on strained $MoTe_2$ ML/EuO heterostructures \cite {Qi}, indicating that achievement of the strained $WS_2$ ML is plausible. More importantly, we found that the lowest energy of the the CBM is located on the $Q$ point under compressive strain, where a large spin splitting is observed [Figs. 3(a) and 4(a)]. For instant, in the case of compressive -2\%, the spin splitting is found to be 0.30 eV. This large spin splitting enables us to allow operation as a spintronics devices at room temperature. Since, large spin splitting is achieved in the CBM by the compressive strain, the $n$-type compressively strained $WS_2$ ML systems for spintronics is expected to be realized. In fact, the $n$-type system of the $WS_2$ ML has recently been experimentally observed \cite {Ovchinnikov}. As such, our findings of the tunable spin splitting under biaxial strain are useful to realizing spintronics applications of $WS_2$ ML system. 

Here, we discuss another possible application of strained $WS_2$ ML by considering the features of the spin textures. Considering the fact that the predicted spin textures in the presence study is merely out-of-plane, suppression of Dyakonov-Perel Mechanism \cite {Zhu,Ohno,Couto} to the spin lifetime is expected to be established, which implies that the carriers have an extended spin lifetime \cite {Zhu, Yuan}. This is supported by the fact that similar mechanism behind long spin lifetime induced by out-of-plane spin orientation has been reported in [110]-oriented zinc-blende quantum well \cite {Ohno,Couto} as well as wurtzite surface systems \cite {Absor2}. Recently, the long-lived nanosecond spin relaxation and spin coherence of electrons in the $WS_2$ ML has been observed \cite {L_Yang}. Therefore, the strained $WS_2$ ML could provide an energy saving spintronics devices.

\section{CONCLUSION}

In conclussion, the spin-orbit induced spin splitting in the CBM of the strained $WS_2$ ML have been investigated by using fully-relativistic first-principles DFT calculations. We found that a substantial spin splitting bands are identified in the CBM, which is effectively controlled and tuned by applying strain. We also found that these spin splitting bands induces spin textures exhibiting fully out-of-plane spin polarization in the opposite direction between the $K$ and $Q$ points and their time reversals. In addition, due to the fully-out-of-plane spin orientations in our calculational results, the long spin lifetime is expected to be achieved due to the suppression of Dyakonov-Perel Mechanism. This is supported by the fact that the long-lived nanosecond spin relaxation and spin coherence of electrons in the $WS_2$ ML has been observed \cite {L_Yang}, suggesting that the strained $WS_2$ ML could provide an energy saving spintronics devices. Recently, the strained $WS_2$ ML system has been extensively studied \cite {David}. Our study clarify that the strain plays an significant role in the SOC of the $WS_2$ ML, which has very important implications in designing future spintronics devices. Given that the $n$-type system of $WS_2$ ML is really achieved in experiment, suggesting that this system is promissing for spintronics applications \cite {Ovchinnikov,Morrish}. 

\begin{acknowledgments}

This work was supported by BOPTN reserach grant funded by Faculty of Mathematics and Natural Sciences, Gadjah Mada University, Indonesia. The computations in this research were performed using the high performance computing facilities (DSDI) at Gadjah Mada University, Indonesia. 

\end{acknowledgments}

\bibliography{WS2_PRB}


\end{document}